\documentclass[a4paper]{article}

\usepackage[T1]{fontenc}
\usepackage{graphicx}
\usepackage{amsmath}
\usepackage{hyperref}
\usepackage{listings}
\usepackage{xcolor}

\usepackage{authblk}

\usepackage{booktabs}
\usepackage{xspace}
\newcommand{\combol}{\textsc{CombOL}\xspace}

\newcommand{\class}[1]{\ensuremath{\mathcal{#1}}}
\newcommand{\param}{\ensuremath{z}}
\newcommand{\params}{\ensuremath{\mathbf{z}}}

\newcommand{\values}{\ensuremath{\mathbf{x}}}
\newcommand{\boltz}{\ensuremath{\mathrm{\Gamma}}}
\newcommand{\op}[1]{\textsc{#1}}  

\newcommand{\fpu}[1]{\overline{#1}}
\newcommand{\fpd}[1]{\underline{#1}}
\newcommand{\precision}{\ensuremath{P}}

\begin{document}
\title{
	\textsc{CombOL}: a Library for Practical Enumeration and Boltzmann Sampling of Combinatorial Classes
}

\author[1,$\ast$]{Casper Asbjørn Eriksen}
\author[2,3,1]{Daniel Merkle}

\affil[1]{
	Department of Mathematics and Computer Science,
	University of Southern Denmark
}
\affil[2]{
	Algorithmic Cheminformatics Group,
	Faculty of Technology, Bielefeld University,
	Germany
}
\affil[3]{
	Center for Biotechnology (CeBiTec),
	Bielefeld University,
	Germany
}
\date{}


%
%
%
\maketitle

\begin{abstract}
	We present \combol (Combinatorial Objects Library), an open-source library for the enumeration and Boltzmann sampling of combinatorial classes.
	Classes can be specified by a concise string syntax, and may depend on an arbitrary number of parameters.
	\combol automatically derives the associated generating functions, enabling the generation of counting sequences and the compilation of Boltzmann samplers
	The library supports exact and approximate-size Boltzmann rejection sampling with automatic parameter tuning to target specific sizes.
	In addition to implementing established methods, \combol contributes a novel early-rejection scheme, as well as guaranteed statistical correctness by dynamically increasing the numerical precision, eliminating bias due to floating-point rounding errors.

	Through the Python interface, sampled structures can be mapped to application-specific objects, enabling direct sampling of domain objects such as graphs, chemical structure representations, or other complex data types.

	\combol is available from PyPI as \texttt{combol} (\href{https://pypi.org/project/combol/}{pypi.org/project/combol}).
	The source code is available at \href{https://gitlab.com/casbjorn/combol}{gitlab.com/casbjorn/combol}.

\end{abstract}

\section{Introduction}
\label{sec:intro}

Analytic combinatorics is a mathematical framework for studying classes of combinatorial objects through their generating functions.
The central idea is to describe such a class symbolically, using set-theoretic constructions such as disjoint unions, Cartesian products, or sequences, and then translate this specification into equations for the associated generating function.
Once the generating function is available, it can be used for a variety of purposes, such as producing the counting sequence, obtaining the asymptotic growth, and uniform random sampling.
As a result, analytic combinatorics is a powerful tool within combinatorics, and random sampling in particular has applications in for example algorithmic engineering, simulation, randomized testing, and the study of large datasets~\cite{applications1,applications2}.

This paper presents \combol, an open-source library for the specification, enumeration, and Boltzmann sampling of combinatorial classes in multiple variables, written in Python and Rust.
Methodologically, \combol builds on the Boltzmann sampling framework~\cite{duchon2004}, the Newton iteration-based oracles and enumeration algorithm~\cite{boltzmannoracle,newtoniteration}, and earlier work on exact uniform generation under floating-point arithmetic~\cite{floatingpoint}.

Several tools with related goals already exist, such as combstruct~\cite{combstruct}, MuPad-CS~\cite{mupadcs}, and Boltzmann Brain~\cite{boltzmannbrain}.
Relative to these predecessors, \combol emphasizes a modern, open-source implementation,  accessibility, and interoperability with Python, allowing direct Boltzmann sampling of domain-specific objects.
In addition to providing efficient implementations of established methods, \combol contributes implementation-level advances: an exact dynamic-precision sampling scheme that prevents bias resulting from numerical rounding, and an early-rejection strategy that improves the efficiency of size-constrained sampling.

The paper is organized as follows.
Section~\ref{sec:usage} introduces the basic notation and gives a brief user-level overview of the specification, enumeration, and sampling in \combol.
Section~\ref{sec:implementation} describes the implementation details, in particular dynamic precision sampling and the early-rejection scheme.
Finally, section~\ref{sec:experimental} summarizes experiments supporting these contributions.

\section{Background and Usage}
\label{sec:usage}
In this section, we briefly introduce the necessary notation from analytic combinatorics and give examples of usage of \combol.
We refer to the standard monograph~\cite{bible} for a general introduction to analytic combinatorics, and to~\cite{samplingreview,duchon2004} for an introduction to Boltzmann sampling.

A combinatorial class (denoted by calligraphic letters e.g. \class{A}) is a class of discrete objects in which each object has a finite integer size described by the variable $z$.
In \emph{multivariate} cases, this size is given by a vector of integers, $\params = [\param_1, \param_2, \cdots, \param_n]$.

The simplest classes are \emph{atoms}, consisting of a single object with a given size.
When the size of that atom is $1$ in one variable and $0$ in all others, the class is denoted by the same symbol as the variable, e.g. $z$.
More complex classes are defined as assemblies starting from atoms, using set-theoretical constructors such as disjoint union ($+$), product ($\times$) and sequence ($\op{Seq}$).
Classes can be defined recursively, potentially in terms of other classes, creating a system of mutually recursive classes.
The canonical example of a combinatorial class is that of binary trees:
$$ \class{B} = z + (z \times \class{B} \times \class{B})
	.
$$
In \combol, a specification can be parsed directly from a concise syntax similar to that of the \emph{combstruct} Maple package.
Translating the previous example:
\begin{lstlisting}
btrees = combol.parse('B = z + (z * B * B)')
\end{lstlisting}

Once a specification is parsed, it is saved as an expression tree describing the mathematical equation, which we call the \emph{symbolic specification tree}, in which leaves represent either atoms ($z$) or recursive class calls ($\class{B}$) and internal nodes represent combinatorial constructions on its children.

A central component of analytic combinatorics is the automatic translation of specifications into generating functions encoding the counting sequence of the class via the so-called \emph{symbolic transfer theorems}.
Given a class $\class{A}$, the corresponding generating function is denoted $A(z)$, where $z$ is a complex control parameter (or a vector of complex control parameters, $\params$, in the general case).
\combol applies the symbolic transfer theorems recursively, producing the generating function for every node in the tree, and can use these to compute counting sequences:
\begin{lstlisting}
btrees.counting_sequence(20)
> [0, 1, 0, 1, 0, 2, 0, 5, 0, 14, 0, 42, 0, 132, 0, 429, 0,
   1430, 0, 4862, 0]
\end{lstlisting}

\subsection{Boltzmann Sampling}
\label{sec:sampling}

A Boltzmann sampler for a class $\class{A}$ is a process $\boltz \class{A}(\values)$ which selects an object at random from $\class{A}$ such that any two objects of the same size have equal probability~\cite{duchon2004}.
If a particular size of object is required, \emph{rejection sampling} can be performed, in which selected objects are discarded if they are not of acceptable size.
Sampling within a tolerance interval of the target size has considerable algorithmic benefits~\cite{duchon2004}.

The Boltzmann sampler is defined recursively for each node in the symbolic specification tree.
For some constructions, such as product ($\times$) and for atomic classes, no choice is made - we call these \emph{deterministic samplers}.
However, a random choice needs to be made for every call to a \emph{non-deterministic} sampler.
For example, the disjoint union sampler $\boltz{(\class{A} + \class{B})}(\values)$ chooses one of its children with probability proportional to the values of the respective generating functions at the chosen control parameters, and the sequence sampler $\boltz{\op{Seq}(\class{A})}(\values)$ needs to choose a cardinality $k$, after which $k$ independent calls to $\boltz{\class{A}}(\values)$ are made.

In order to calculate the probabilities for every outcome, numerical values for the generating functions are necessary.
In general, it is not possible to compute the closed form of generating functions (e.g. quinary trees due to the Abel-Ruffini theorem), so numerical methods are necessary to approximate the values up to a given precision (this procedure is called the \emph{oracle}).
Given such values for every class, the generating function values for each node in the specification tree are calculated recursively, after which a tree of samplers can be compiled based on these values.

In \combol, sampling a class requires first compiling a sampler specific to a given set of values of the control parameters.
\begin{lstlisting}
sampler = btrees.sampler({'z': 0.2})
sampler.sample(n = 1000)
\end{lstlisting}

Alternatively, control parameters can be automatically selected (`\emph{tuned}') to optimize for a given size of structure, or to produce samplers at singular values (using the implementation of~\cite{param_tuning}).
The following samples binary trees with a size within a 5\,\% tolerance of 1000 nodes.
\begin{lstlisting}
btrees.sample(
    n = 1000,
    target_size: {'z': 1000},
    tolerance = 0.05
)
\end{lstlisting}

\subsection{Integration}
In addition to generating native symbolic structures, \combol allows integration of this sampling procedure into other workflows.
By defining the atoms and how operators such as \op{Product} and \op{Sequence} act, the user can then use \combol to directly generate objects of arbitrary Python types, such as text-based representations, or domain-specific types such as chemical or biological structures.

\begin{figure}
	\centering
	\includegraphics[width=\textwidth]{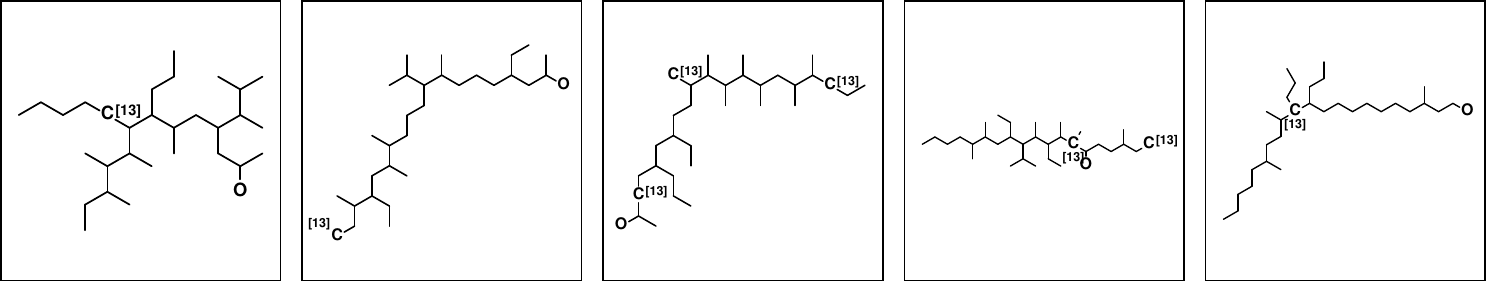}
	\caption{
		Five molecular structures, generated by \combol using the script included in Appendix~\ref{app:example}.
	}
	\label{fig:mols}
\end{figure}

Appendix~\ref{app:example} contains an example of such an integration, in which we consider the class of acyclic hydrocarbon molecules with some carbon atoms replaced by the isotope 13C.
Say we want to sample structures with a realistic distribution of isotopes, for instance as test instances for an analytical chemistry tool.
Using \combol, we specify a multivariate class with one variable counting carbon atoms in general, and another counting the 13C isotopes.
We can then perform multivariate boltzmann sampling, with control parameters automatically being selected according to a specified target size which models a realistic distribution of isotopes while automatically accounting for stereochemical symmetry.
Fig.~\ref{fig:mols} shows an example of the output.

\section{Implementation}
\label{sec:implementation}
\combol is implemented primarily in Rust, with an interface and high-level functionality such as specification parsing provided through Python bindings.
The mathematical backend is the Rust-based computer algebra library \emph{Symbolica} and numerical library \emph{Rug}.

For producing the counting sequence, \combol implements the Newton iteration algorithm of~\cite{newtoniteration}, in which the number of correct terms increases quadratically for each iteration.
The same method is used for the Boltzmann oracle, producing generating function values instead of counting sequences.

%
%
\subsection{Boltzmann Sampling}
\label{sec:impl_sampling}
The sampling procedure in \combol is developed with the aim of being as fast as possible while also delivering bias-free sampling.
This was achieved by dynamic-precision floating-point calculations, using hardware-accelerated floating point calculations when possible and increasing the precision whenever required to prevent uncertainty.

\subsubsection*{Sampler Compilation}
Before sampling, we compute the probabilities used in each non-deterministic node given parameter values $\values$, and the numerical precision $\precision$, essentially \emph{compiling} a tree of Boltzmann samplers specific to these parameters.
For a non-deterministic node, let $p_1, p_2, \cdots$ represent the probability of each of the (possibly infinite) outcomes.
These probabilities are based on the GF values of the child nodes, of which conservative lower and upper bounds are calculated using error-propagating floating-point calculations.
From these, we obtain bounds on the probabilities, denoted by $l_n$ and $u_n$, $l_n \leq p_n \leq u_n$, such that $l_n$ and $u_n$ are representable in precision $\precision$.

\begin{figure}
	\centering
	\includegraphics[scale=0.8]{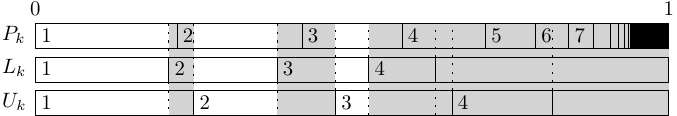}
	\caption{
		Illustration of the true cumulative probabilities, $P_k$, and the lower and upper bounds, $L_k$ and $U_k$.
		Grey regions represent values of $r$ for which the result is ambiguous.
		In this (exaggerated) case, terms beyond 4 are always ambiguous and do not need to be computed.
	}
	\label{fig:bounds}
\end{figure}

If the values were known exactly, a random choice could be made by choosing a random value $r \in [0, 1]$ and comparing against the cumulative probabilities $P_k = \sum_1^k p_k$ to determine the outcome.
We replicate this behaviour under floating-point arithmetic by computing the rounded bounds on the cumulative probabilities $L_k$ and $U_k$:
$$ L_k = \fpd{ \left( \fpd{ \left( \fpd{l_1 + l_2} \right) + l_3} \right) + \cdots + l_k}
	\text{\quad and \quad}
	U_k = \fpu{ \left( \fpu{ \left( \fpu{u_1 + u_2} \right) + u_3} \right) + \cdots + u_k}
	,
$$
where $\fpd{d}$ (resp. $\fpu{d}$) represent rounding $d$ to the previous (next) representable floating-point value.
Using these bounds, we proceed iteratively: given a random number $r$ at step $k$, if $r < L_k$, outcome $k$ can be returned with certainty, and if $r > U_k$, we can proceed to $k+1$.
However, if $L_k \leq r \leq U_k$, the result is ambiguous, and we need to narrow the bounds by re-compiling the sampler at a greater precision.
Fig.~\ref{fig:bounds} shows an exaggerated example of these outcomes.
This procedure implies a natural point at which we can stop computing terms for unbounded cardinality constructors such as \op{Seq}: When $U_{k} \geq L_{k+1}$, the first condition can never be fulfilled.
When the random value $r$ is within an ambiguous region, sampling needs to be restarted from the same state after re-compiling the sampler with increased precision $\precision'$.
This is achieved by re-creating the state of the sampler, choosing a random value $r'$ such that $r = r'$ in precision $\precision$, and then continuing the process, potentially re-compiling once again if $r'$ is also within an ambiguous region.

\subsubsection*{Optimizations}
\combol implements several novel optimizations to the classical Boltzmann sampling method.
Rather than constructing the output object immediately, the sampler first records the sequence of random choices.
This has the advantage of not spending cycles assembling structures when the attempt may be rejected, and furthermore enables easily re-creating the state of the sampler after it is re-compiled to increase precision.
A structure is then actually constructed only when a sample is accepted - this happens within the Python interface, where the specification tree is traversed according to the choices, and user-supplied methods are called to construct the appropriate object.
Additionally, the standard recursive sampling procedure is replaced by an iterative process, keeping a \emph{sampler stack} of samplers to be evaluated, enabling the early rejection-scheme.

It is desirable to reject a sampling attempt as early as possible as many rejections are expected for each accepted sample, in general.
Instead of rejecting only when the partial object already exceeds the admissible upper size bound, \combol tracks the size contribution of the sampler stack.
At each step, the sizes of the already chosen atoms together with the cumulative lower and upper bounds of the pending stack yields exact bounds on the possible final size of the object.
Sampling can therefore be aborted as soon as acceptance becomes impossible.

\section{Experimental results}
\label{sec:experimental}
We briefly evaluate the two implementation-level contributions of \combol: exact dynamic-precision sampling and early rejection.

\subsubsection*{Uniformity}
To assess the uniformity of the Boltzmann sampler, we considered the distribution of sampled objects for various sizes, and compared the empirical frequencies against the uniform distribution using Pearson's $\chi^2$ test for several classes and sizes.
The results are summarized in Table~\ref{tab:uniform}.

\begin{table}
	\centering
	\begin{tabular}{@{}lrrrlrrrlrrr@{}}
		\toprule
		                                               &
		\multicolumn{3}{c}{Binary trees, $\class{B}$}  &
		                                               &
		\multicolumn{3}{c}{General trees, $\class{T}$} &
		                                               &
		\multicolumn{3}{c}{UB trees, $\class{UB}$}                                                                         \\ \midrule
		                                               &
		\multicolumn{1}{c}{\#}                         &
		\multicolumn{1}{c}{$\chi^2$}                   &
		\multicolumn{1}{c}{$p$}                        &
		                                               &
		\multicolumn{1}{c}{\#}                         &
		\multicolumn{1}{c}{$\chi^2$}                   &
		\multicolumn{1}{c}{$p$}                        &
		                                               &
		\multicolumn{1}{c}{\#}                         &
		\multicolumn{1}{c}{$\chi^2$}                   &
		\multicolumn{1}{c}{$p$}                                                                                            \\ \cmidrule(lr){2-4} \cmidrule(lr){6-8} \cmidrule(l){10-12}
		\textbf{3}                                     & 2   & 0.231 & 0.63 &  & 2   & 0.041 & 0.84 &  & 2  & 4.401 & 0.36 \\
		\textbf{4}                                     & 5,  & 1.234 & 0.87 &  & 5   & 4.752 & 0.32 &  & 4  & 2.021 & 0.57 \\
		\textbf{5}                                     & 14  & 15.61 & 0.27 &  & 14  & 11.64 & 0.56 &  & 9  & 9.480 & 0.30 \\
		\textbf{6}                                     & 42  & 44.27 & 0.34 &  & 42  & 35.72 & 0.70 &  & 21 & 12.20 & 0.91 \\
		\textbf{7}                                     & 132 & 108.4 & 0.93 &  & 132 & 145.3 & 0.19 &  & 51 & 60.71 & 0.14 \\ \bottomrule
	\end{tabular}
	\caption{
		For 100\,000\,000 samples of the classes $\class{B} = z + (\class{B} \times \class{B})$, $\class {T} = z + (z \times \op{SEQ}(\class{T})$, and $\class{UB} = z + (z \times \class{UB}) + (z \times \class{UB} \times \class{UB})$, the number of distinct structures (\#), the $\chi^2$-value compared to the uniform distribution, and the $p$-value of the test.
	}
	\label{tab:uniform}
\end{table}

In all cases, we observe no statistically significant deviation from the expected uniform distribution ($p > 0.05$).
It is notable that the precision was rarely increased above the precision of a double-precision floating-point (53 bits) in practical tests, and that tests simulating the precision of double-precision floating-point numbers yielded similar results.
Therefore, for most practical purposes, the exactness guarantee has little practical consequence, as in order to produce a statistically significant bias due to precision errors, it would require an extremely large sample size or specifically tailored cases.

\subsubsection*{Early Rejection}
We also quantified the impact of our early-rejection scheme against a traditional rejection scheme.
The impact depends on the particular class being sampled, the parameter values, and the admissible size range, so we do not make any general claims.
As an example, in the case of binary trees of size [40, 60] sampled at $z = 0.48$, early rejection improved the runtime by a factor of $2.44$ on average (95\,\% CI: $2.01$-$2.94$), indicating that this scheme can provide a meaningful practical speedup in size-constrained sampling in at least some cases.

\section{Conclusion}
\label{sec:conc}

With \combol, we aim to make methods from analytic combinatorics accessible as practical mathematical software.
It combines symbolic specification, enumeration, parameter tuning, and exact Boltzmann sampling in a modern open-source library with Python interoperability.

The current iteration serves as a starting point for further development.
Future work includes extending support to labeled classes and more combinations of constructors and constraints, supporting asymptotic analysis where possible, and exploring additional combinatorial tasks such as ranking and unranking.

%
%
\bibliographystyle{splncs04}
\bibliography{refs}

%
%
\newpage
\section*{Appendices}
\begin{appendix}
	\section{Example}
	\label{app:example}

	The following listing defines a combinatorial class, then defines a builder representing the combinatorial bijection between the class and the class of acyclic hydrocarbons.
	The output of this code can be seen in Fig.~\ref{fig:mols}.

	\begin{lstlisting}
from combol import *
import rdkit.Chem as rdkit

mol_class = parse(
    'M = C * (C * M) + (C * M * M) + (C * M * M * M)',
    'C = c + c13',
    'c = atom(z: 1)',
    'c13 = atom(z: 1, u: 1)'
)

# Define the builder
class MolBuilder(StructureBuilder[rdkit.Mol, rdkit.Mol]):
    def atom(self, atom: Atom) -> rdkit.Mol:
        # Generate an atom connected to a '*' atom,
        # representing the 'attachment point'.
        match atom.key:
            case 'c':
                return rdkit.MolFromSmiles('C*')
            case 'c13':
                return rdkit.MolFromSmiles('[13C]*')

    def product(self, constructor, children) -> rdkit.Mol:
        # First item is parent, rest are children
        mol = rdkit.EditableMol(
            rdkit.CombineMols(parent, *children[1:])
        )
        for atom in mol.GetMol().GetAtoms():
            if atom.GetAtomicNum() == 0:
                # Replace bond atom by bond
                neighbour = atom.GetNeighbors()[0].GetIdx()
                if neighbour != 0:
                    mol.AddBond(0, neighbour, rdkit.BondType.SINGLE)
                    mol.RemoveAtom(atom.GetIdx())
        return mol

    def finalize(self, obj: rdkit.Mol) -> rdkit.Mol:
        # Append OH group to the root
        # (replace the remaining dummy atom with O)
        rw = rdkit.RWMol(obj.GetMol())
        for atom in rw.GetAtoms():
            if atom.GetAtomicNum() == 0:
                atom.SetAtomicNum(8)

        mol = rw.GetMol()
        rdkit.SanitizeMol(mol)
        return mol

# Sample
c13_abundance = 0.0103
mols: list[rdk.Mol] = mol_class.sample(
    n = 10,
    target_size = {'z': 50, 'u': 50 * c13_abundance},
    builder = MolBuilder()
)



	\end{lstlisting}
\end{appendix}

\end{document}